\documentclass[article]{jss}
\makeatletter
\@nojsstrue
\makeatother
\usepackage{amsmath}
\usepackage{color}
\definecolor{darkblue}{rgb}{0.0,0.0,0.75}
\definecolor{distrCol}{rgb}{0.0,0.4,0.4}
\usepackage{amssymb}%
\usepackage{amsthm}%

\newcommand{\Ss}{\scriptstyle}  

\newcommand{\Jc}{\mathop{\bf\rm{{}I{}}}\nolimits}
\newcommand{\N}{\mathbb N}
\newcommand{\R}{\mathbb R}
\newcommand{\C}{\mathbb C}
\newcommand{\B}{\mathbb B}
\newcommand{\Z}{\mathbb Z}
\newcommand{\sZ}{{\Ss \mathbb Z}}
\newcommand{\Tfrac}[2]{\textstyle\frac{#1}{#2}}
\newcommand{\hreff}[1]{\href{#1}{\footnotesize\url{#1}}}

\newtheorem{Thm}{Theorem}
\newtheorem{Prop}[Thm]{Proposition}

\newtheorem{Rem}[Thm]{Remark}
\newtheorem{Exa}[Thm]{Example}

\newtheorem{algo}[Thm]{Algorithm}
\numberwithin{equation}{section}
\numberwithin{Thm}{section}

\author{Peter Ruckdeschel\\TU Kaiserslautern and\\Fraunhofer ITWM Kaiserslautern\And
        Matthias Kohl\\Furtwangen University}
\title{General Purpose Convolution Algorithm in S4-Classes by means of FFT}
\Plainauthor{Peter Ruckdeschel, Matthias Kohl} 
\Plaintitle{General Purpose Convolution Algorithm in S4-Classes by means of FFT} 
\Shorttitle{General Purpose FFT Convolution Algorithm} 

\Abstract{
Object orientation provides a flexible framework for the implementation
of the convolution of arbitrary distributions of real-valued random variables.

We discuss an algorithm which is based on the discrete Fourier transformation
(DFT) and its fast computability via the fast Fourier transformation (FFT). It directly
applies to lattice-supported distributions. In the case of continuous
distributions an additional discretization to a linear lattice is necessary
and the resulting lattice-supported distributions are suitably smoothed after
convolution.

We compare our algorithm to other approaches aiming at a similar generality
as to accuracy and speed. In situations where the exact results are known,
several checks confirm a high accuracy of the proposed algorithm which is also
illustrated at approximations of non-central $\chi^2$-distributions.

By means of object orientation this default algorithm can be overloaded
by more specific algorithms where possible, in particular where explicit
convolution formulae are available.

Our focus is on \proglang{R} package \pkg{distr} which implements
this approach, overloading operator ``$+$''for convolution; based
on this convolution, we define a whole arithmetics of
mathematical operations acting on distribution objects,
comprising, among others, operators \code{+}, \code{-},
\code{*}, \code{/}, and {\tt \textasciicircum}.

}
\Keywords{probability distributions, FFT, convolution,
random variables, \proglang{S4}~classes, \proglang{S4}~methods}
\Plainkeywords{probability distributions, convolution, FFT,
random variables, S4 classes, S4 methods} 

\Address{
  Peter Ruckdeschel\\
  TU Kaiserslautern, Dept. of Mathematics\\
  P.O.Box 3049\\
  67653 Kaiserslautern, Germany\\
  and\\
  Fraunhofer-ITWM, Dept. Financial Mathematics\\
  Fraunhofer-Platz 1\\
  67663 Kaiserslautern, Germany\\
  E-mail: \email{Peter.Ruckdeschel@itwm.fraunhofer.de}\\
  \bigskip\\
  Matthias Kohl\\
  Department of Mechanical and Process Engineering\\
  Furtwangen University\\
  Jakob-Kienzle-Str.\ 17\\
  78054 Villingen-Schwenningen, Germany\\
  E-mail: \email{Matthias.Kohl@stamats.de}
}

\begin{document}
%
\section{Motivation}
Convolution of (probability) distributions is a standard problem in statistics.
For its implementation the Fast Fourier Transformation (FFT) has been
common practice 
ever since the appearance of \citet{Co:Tu:65}.

Combined with an object oriented programming (OOP) approach, this technique gets
even more attractive: We may use it as a default algorithm in situations
where no better alternative is known, while in special cases
as e.g., those of normal or Poisson random variables, where convolution reduces
to transforming the corresponding parameters, a dispatching mechanism realizes
this and replaces the general method by a particular (possibly exact) one.
The user does not have to interfere with the dispatching mechanism, but
is rather provided with one single function/binary operator
for the task of convolution.

We discuss this approach within the \proglang{R} project (cf.\ \citet{RMANUAL})
where it is implemented in package \pkg{distr}, available on
\href{http://cran.r-project.org/mirrors.html}{\tt CRAN}.
Package \pkg{distr} provides classes for probability
distributions within the \proglang{S4} OOP-concept of \proglang{R};
see \citet{RKSC:06,distr}.

In this context, convolution is the workhorse for setting up
a whole arithmetics of mathematical operations acting on distribution objects,
comprising, among others, operators \code{+}, \code{-}, \code{*}, \code{/}, and
{\tt \textasciicircum}.
In this arithmetics, we identify distributions with corresponding (independent)
random variables: If  \code{X1} and \code{X2} are corresponding distribution variables,
\code{X1+X2} will produce the distribution of the sum of respective
(independent) random variables, i.e., their convolution. Technically,
speaking in terms of programming, we have overloaded the operator
``\code{+}'' for univariate distributions.

Convolution itself is computed according to the actual classes
of the operands, with particular (exact) methods for e.g., normal or
Poisson distributions.
\begin{Schunk}
\begin{Sinput}
R> library("distr")
R> N1 <- Norm(mean = 1, sd = 2)
R> N2 <- Norm(mean = -2, sd = 1)
R> N1 + N2
\end{Sinput}
\begin{Soutput}
Distribution Object of Class: Norm
 mean: -1
 sd: 2.23606797749979
\end{Soutput}
\end{Schunk}
In the default method distributions are discretized to lattice form and
the Discrete Fourier Transformation (DFT) is applied. Thus, our general-purpose
algorithm needs no assumptions like Lebesgue densities.
\begin{Schunk}
\begin{Sinput}
R> U1 <- Unif(Min = 0, Max = 1)
R> U3 <- convpow(U1, N = 3)
R> plot(U3, cex.inner = 1,
+       inner = c("density", "cdf", "quantile function"))
\end{Sinput}
\end{Schunk}
\begin{figure}[!ht]
  \begin{center}
    \includegraphics[width=12cm]{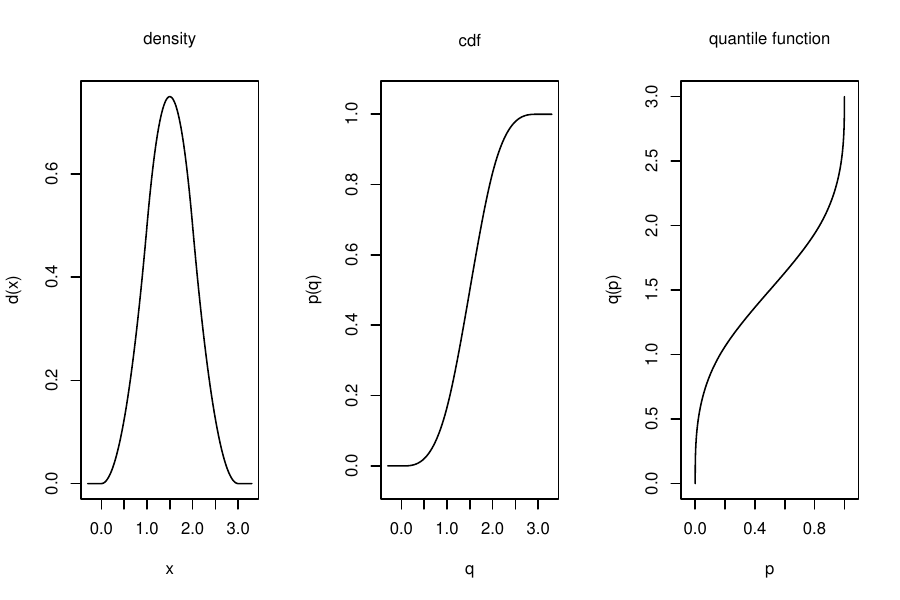}%
    \caption{\label{figmyFam}Plot of 3-fold convolution of a \code{Unif(0,1)} object.}
  \end{center}
\end{figure}%

While all our applied techniques are not novel in themselves,
and much of the infrastructure (FFT in particular) has already been available in
\proglang{R} for long, the combination as present in our approach is unique.
Neither in core \proglang{R} nor in any other contributed add-on package
available on the standard repositories, i.e.,
\href{http://cran.r-project.org/mirrors.html}{\tt CRAN},
\href{http://www.bioconductor.org/}{Bioconductor}, or
\href{https://www.rmetrics.org/}{Rmetrics}, there
is a similarly general approach: We provide a \texttt{"+"} (aka convolution)
operator applying to [almost] \textit{arbitrary} univariate distributions,
no matter whether discrete or continuous;  more
specifically we cover every distribution that is representable as a convex
combination of an absolutely continuous distribution and a discrete distribution.
In addition, the return value of this \texttt{"+"} operator is again a
distribution object, i.e., consisting not only of either a cumulative
distribution function (cdf) or a density/probability
function, but automatically of all four constitutive functions, i.e., cdf,
density, quantile function, and  random number generator.
Accuracy of our default methods can be controlled through global options,
see \code{?distroptions}. Just to illustrate our point, we take up the initial
example and compute the $1/3$-quantile of the convolution
${\cal N}(1,2)\ast {\rm unif}(0,1)^{\ast 3} \ast {\rm Poisson}(1)$, as well as
evaluate its density at the vector $(0.5,0.8)$
\begin{Schunk}
\begin{Sinput}
R> P <- Pois(lambda=1)
R> D <- N1 + U3 + P
R> q(D)(1/3)
\end{Sinput}
\begin{Soutput}
[1] 2.490786
\end{Soutput}
\begin{Sinput}
R> d(D)(c(0.5,0.8))
\end{Sinput}
\begin{Soutput}
[1] 0.07526700 0.08894159
\end{Soutput}
\end{Schunk}

The approach is not restricted to academic purposes: the results are sufficiently
accurate to be used in practice in many circumstances: Be it quite general
compound distribution models as relevant in actuarial sciences, be it very flexible
model fitting techniques as described in detail in \citet{Ko:Ru:10a},
or be it very general robustification techniques as in packages \pkg{RobAStBase},
\citep{RobAStBase},
\pkg{ROptEst}, \citep{ROptEst}, and specialized to Biostat applications in
\pkg{RobLoxBioC}, \citep{RobLoxBioC}, compare \citet{Koh:04di} and \citet{K:D:10}.

When interest lies in multiple
convolutions (of identical summand distributions) we provide a function
\code{convpow} to quickly and reliably compute convolution powers; in particular
sample size then is not an issue. Otherwise, i.e., for non-identically
distributed summands, you either have to appeal to asymptotics in some way
or do it summation by summation. We can say though, that our approach works
reliably to up to 40 (non-)iid summands. In each case, we automatically
provide respective quantile functions which are of particular interest
in actuarial sciences and risk management.

\indent Our paper is organized as follows:

In Section~\ref{oopsec}, we discuss how an object oriented framework
could enhance implementations of both probability distributions in
general and convolution algorithms in particular. To this end, we
sketch our implementation of distribution classes in \proglang{R}
package \pkg{distr}. We also briefly discuss the dispatching decisions
involved when a new object of a distribution class is generated by
convolution. In Section~\ref{sect4}, we present the general purpose
FFT-Algorithm and some ramifications. Some forerunners in this direction
and connections to other approaches are discussed in
Section~\ref{otherapp}.
In Section~\ref{ap.fft.check} we present checks for the accuracy and
computational efficiency of our algorithm. At the end of this paper
we provide some conclusions in Section~\ref{conclusion}
\section{OOP for probability distributions and convolution}\label{oopsec}
\subsection{OOP for probability distributions}\label{OOPdistr}
There is a huge amount of software available providing functionality
for the treatment of probability distributions. In this paper we will mainly
focus on \proglang{S}, more specifically, on its Open Source implementation
\proglang{R}, but of course the considerations also apply for other extensible
software like \proglang{XploRe}, \proglang{Gauss}, \proglang{Simula},
\proglang{SAS} or \proglang{MATLAB}. 
All these packages provide standard distributions,
like normal, exponential, uniform, Poisson just to name a few.\\
There are limitations, however: You can only use distributions which either
are already implemented in the package or in some add-on library, or distributions
for which you yourself have provided an implementation. Automatic
generation of new distributions is left out in general.

In many natural settings you want to formulate algorithms once for
all distributions, so you should be able to treat the actual distribution,
say \code{D}, as argument to some function. This requires particular data types
for distributions. Going ahead in this direction, you may wish to formulate
statements involving the expectation or variance of functions of
random variables as you are used to in Mathematics; i.e., no matter
if the expectation involves a finite sum, a sum of infinite summands,
or a (Lebesgue) integral.
This idea is particularly well-suited for OOP, as
described in \citet{Boo:95}, with its paradigms ``inheritance'' and ``method overloading''.\\
In the OOP concept, we could let a dispatching mechanism decide which method to
choose at run-time. In particular, the result of such an algorithm may be a new
distribution, as in our convolution case.

%
In his \proglang{Java} MCMC-simulation package {\tt HYDRA}, \citet{HYDRA}
heads for a similar OOP approach. Under
\href{http://statdistlib.sourceforge.net/}{\small \tt http://statdistlib.sourceforge.net/},
the author provides a set of \proglang{Java} classes representing common statistical
distributions, porting the \proglang{C}-code underlying the \proglang{R} implementation.
But, quoting the author himself from the cited web-page, ``[o]ther than grouping the {\tt PDF},
{\tt CDF}, etc into a single class for each distribution, the files don't (yet)
make much use of OO design.''
%
\subsection[OOP in S: The S4-class concept]{OOP in \proglang{S}: The \proglang{S4}-class concept}\label{OOPc}
%
In base \proglang{R}, OOP is realized in the \proglang{S3}-class concept as
introduced in \citet{Cham:93a,Cham:93b}, and by its successor, the
\proglang{S4}-class concept, as developed in \citet{Cham:98,Chamb:99} and
described in detail in \citet{Cha:08}. We work with the \proglang{S4}-class
concept.\\
Using the terminology of \citet{Beng:03}, this concept is intended to be {\it FOOP\/}
(function-object-oriented programming) style, in contrast to {\it COOP\/}
(class-object-oriented programming) style, which is the intended style in \proglang{C++},
for example.\\
In COOP style, methods providing access to or manipulation of an object are part of the object,
while in FOOP style, they are not, but rather belong to so-called
{\it generic functions\/} which are abstract functions allowing for arguments of
varying type/class.
A dispatching mechanism then decides on run-time which method best fits
the {\it signature\/} of the function, that is, the types/classes of (a certain subset of)
its arguments. In \proglang{C++}, ``overloaded functions'' in
the sense of \citet[Section 4.6.6]{Stro:92} come next to this concept.
\\
FOOP style has some advantages for functions like ``\code{+}'' having a natural meaning
for many operand types/classes as in our convolution case.
It also helps collaborative programming, as not every programmer
providing functionality for some class has to interfere into the original class definition.
In addition, as \proglang{S} respectively, \proglang{R} is an interpreted language, a method
incorporated in a \proglang{S4}-class definition would not simply be a pointer but rather
the whole function definition and environment. Hence, the COOP-style paradigm in (standard)
\proglang{R}  entails arguable draw-backs and hence is not generally advisable within
the \proglang{S4}-class system. Since \proglang{R} version {\tt 2.12.0},
this has been overcome to some extent, however, with the introduction of
reference classes.

Since its introduction to \proglang{R}, the \proglang{S4}-class concept
has allowed COOP style, that is, members (or {\it slots\/} in \proglang{S4}-lingo)
have always been permitted to be functions,  but we may say that
use of functional slots in \proglang{S4} is not standard, which may be judged
against a thread on the \proglang{R} mailing list \texttt{r-devel}
on \url{http://tolstoy.newcastle.edu.au/R/devel/04a/0185.html}.
Use of functional slots has been extensively used in \citeauthor{Beng:03}'s
\citeyearpar{Beng:03} \pkg{R.oo} package where the author circumvents the
above-mentioned problems by a non-standard {\it call-by-reference\/} semantic.

For our distribution classes, we, too, use the possibility for function-type members,
albeit only in a very limited way, and not extending the standard \proglang{S4}
system in any respect. Still, others have suggested to rather follow the
\proglang{S4}-generic way for slots \texttt{r,d,p,q},
 which however, in our opinion, would lead to many class definitions
[a new one generated at each call to the convolution operation]
instead of only few class definitions as in our design.

%
\subsection[Implementation of distribution classes]%
{Implementation of distribution classes within the %
\proglang{S4}-class concept}
%
In \proglang{S}/\proglang{R}, any distribution is given through four functions:
{\tt r}, generating pseudo-random numbers according to that distribution,
{\tt d}, the density or probability function/counting density,
{\tt p}, the cdf, and
{\tt q}, the quantile function.
This is also reflected in the naming convention
{\tt [prefix]<name>} where {\tt [prefix]} stands for
{\tt r}, {\tt d}, {\tt p}, or {\tt q}  and {\tt <name>} is the (abbreviated) name
of the distribution.\\
We call these functions {\em constitutive} as we regard them as integral part
of a distribution object, and hence realize them as members (slots) of our distribution
classes even though this causes some ``code weight'' for the corresponding objects.
A real benefit of this approach is grouping of routines which represent
one distribution instead of having separate functions
\code{rnorm}, \code{dnorm}, \code{pnorm}, and \code{qnorm} which otherwise
are only  connected by gentleman's agreement / naming convention.

Consistency may become an issue then, of course: We cannot exclude the possibility
that someone (inadvertedly) puts together inadequate {\tt r}, {\tt d}, {\tt p}, or
{\tt q} slots, manipulating the slots by assignments of the like \code{a@b <- 4}.
This is not the intended way to generate distribution objects, though.
We do have generating functions for this purpose, the return values
of which are consistent; the same goes for automatically generated
distributions arising as return values from arithmetic operations.
In addition, we do provide a certain level of consistency, following
\citet{OOPGent} and providing corresponding accessor- and replacement
functions for each of the slots. We strongly discourage the use of the
\code{@}-operator to modify or even access slots {\tt r}, {\tt d}, {\tt p},
and {\tt q} and explicitly have mentioned this in
\citet[section~9 and Example~13.7]{RMandistr} at least since 2005.
\\
Another justification for this approach can be given by considering convolution:
Assume we would like to automatically generate the constitutive functions for the law
of expressions like \code{X+Y} for objects \code{X} and \code{Y} of some distribution
class. Following the FOOP paradigm the function \code{cdf} to compute the cdf would
not be part of the class but some method of a corresponding generic function.
Then, as the constitutive functions vary from distribution to distribution and the
dispatching mechanism makes its decision which method to use for \code{cdf} based on
the signature, we would have to derive a new method for \code{cdf} for every
(new) distribution class and would in particular need a new class for every newly
generated distribution. That is, very soon the dispatching mechanism would have to decide
between lots of different signatures. In contrast, when \code{cdf} is a member of a class,
dispatching is not necessary and calculations are more efficient. This efficiency
is not obtained by extracting the, say, cdf as a functional slot,
instead of getting it from dispatch after a quick look-up in a hash table,
but rather by the necessity to have a sufficiently general class
for the return value of convolution of arbitrary distributions:
As a rule, the convolution of two arbitrary distributions $f$ and $g$ will
generate a new distribution $f\ast g$ for which there has not been an
implementation before.
So in order to have access to $f\ast g$ in FOOP manor, you either
have to compute cdf or density or quantile function ``on the fly''
for each evaluation or you have to generate a new S4 class
and a hash table to re-find the particular cdf of $f\ast g$ when calling
something like \texttt{cdf(conv(f,g))} or you have to limit the class
of admitted operands (arguments) of \texttt{conv()}, such that the
result object is again a member of a (possibly parametric)
set of distribution functions.

In fact, \proglang{R} package \pkg{actuar}, \citet{D:G:P:08},
pursues the FOOP approach just sketched in their function \code{aggregateDist}.
To escape the possible multitude of new distribution classes, the authors
restrict themselves to particular probability distributions; i.e., the
``the (a, b, 0) or (a, b, 1) families of distributions'' (see cited reference
for their definition).
Doing so, they can offer alternatives to compute the convolution (see help
to \code{aggregateDist}).

Their approach and ours do combine well though: Our extension package \pkg{distrEx}
even depends on package \pkg{actuar}, using some of the additional root distributions
provided there; these distributions are implemented efficiently as sets of
functions interfacing to \proglang{C}, and their names follow
the above-mentioned {\tt [prefix]<name>} paradigm.

%
\subsection[Convolution as a particular method in "distr"]{Convolution as a particular method in \pkg{distr}}
%
Contrary to the {\tt r}, {\tt d}, {\tt p}, and {\tt q} functions just discussed,
the computation of convolutions ideally fits in the FOOP-setup where
method dispatching works as follows:
\\
In the case that there are better algorithms or even exact convolution formulae
for the given signature, as for independent variables
distributed according to ${\rm Bin}(n_i,p)$, $i=1,2$, or ${\rm Poisson}(\lambda_i)$
or ${\cal N}(\mu_i,\sigma_i^2)$ etc., the dispatching mechanism for \proglang{S4}-classes
will realize that, will use the best matching existing ``\code{+}''-method and
will generate a new object of the corresponding class. However, this case is
exceptional. Hence, we do not have to dispatch among too many methods.\\
As our object oriented framework allows to override the default procedure easily by
more specialized algorithms by method dispatch, the focus of our default algorithm,
Algorithm~\ref{ap.fft.algo.algo}, is not to provide the most refined techniques
to achieve high accuracy but rather to be applicable in a most general setting.
This default algorithm is based on FFT and will be described
in detail in the next section. It originally applies to distribution objects
of class \code{LatticeDistribution}. A lattice distribution is a discrete distribution
whose support is a lattice of the form $a_0+iw$, $a_0\in\R$,
$w\in\R\setminus\{0\}$ with $i\in\N_0$ (or $\{0,1,\ldots,n\}$, $n\in\N$). In our implementation
this class is a subclass of class \code{DiscreteDistribution} which in addition to its
respective mother class \code{UnivariateDistribution} has an extra slot \code{support}, a numerical
vector containing the support (if finite, and else a truncated version carrying
more than $1-\varepsilon$ mass). Besides \code{DiscreteDistribution}, class
\code{UnivariateDistribution} has subclasses \code{AbscontDistribution} for
absolutely continuous distributions, i.e., distributions with a (Lebesgue) density,
and \code{UnivarLebDecDistribution} for a distribution in Lebesgue decomposed form,
i.e., a mixture of an absolutely continuous part and a discrete part. Such distributions
e.g., arise from truncation operations, or when a discrete distribution (with point mass at $\{0\}$)
is multiplied with a(n) (absolutely) continuous one.\\
Our FFT-based algorithm starts with two lattice distributions with compatible lattices;
i.e., we assume that the support of the resulting convolved distribution has length
strictly smaller than the product of the lengths of the supports of the operands.
For discrete distributions, we check whether they can be cast to lattice distributions
with compatible lattices. If one operand is absolutely continuous, the other one discrete,
we proceed by ``direct computation''. If both operands are absolutely continuous, as
described in Algorithm~\ref{ap.fft.algo.algo}, we first discretize them to lattice
distributions with same width $w$. The cdfs $F_1$ and $F_2$ used in this algorithm will
be obtained from the corresponding {\tt p}-slots.
For objects of class \code{UnivarLebDecDistribution}, we proceed component-wise.\\
Slots {\tt p} and {\tt d} of the resulting new object are then filled by
Algorithm~\ref{ap.fft.algo.algo}, described in detail in the next section.
More precisely we will use variants of this algorithm for the absolutely continuous
and the discrete/lattice case, respectively.\\
Slot {\tt r} of the new object consists in simply simulating pairs of variables by means of
the {\tt r} slots of the convolutional summands and then summing these pairs.
Slot {\tt q} is obtained by numerical inversion: For a continuous approximation
of the quantile function we evaluate the function in slot {\tt p} on an $x$-grid,
exchange $x$- and $y$-axis and interpolate linearly between the grid points, for
discrete distributions \code{D} we start with the vector \code{pvec <- p(D)(support(D))}
and search for the support-point belonging to the largest member of \code{pvec} smaller
than or equal to the argument of \code{q}.
%
\subsection[General arithmetics of distributions in distr]%
{General arithmetics of distributions in \pkg{distr}}
%
An important consequence of our approach of implementing distributions as
classes is that this enables us to implement a fairly complete and accurate
arithmetics acting on distributions respectively on random variables with
corresponding distributions.\\
The first observation to be made is that the image distribution
of affine linear transformations can be explicitly spelt out for each of
the slots {\tt r}, {\tt d}, {\tt p}, and {\tt q}. Hence, if \code{X} and \code{Y}
are both univariate distributions, we define \code{X-Y} to mean the convolution
of \code{X} and \code{-Y}. For distributions with support contained in $(0,\infty)$,
also multiplication is easy: as $\log$ and $\exp$ are strictly monotone
and differentiable transformations, the respective image distributions may
also be spelt out explicitly, for each of the slots {\tt r}, {\tt d}, {\tt p},
and {\tt q}, and the {\tt X*Y=exp(log(X)+log(Y))}. Splitting up the support of
a distribution into positive, negative, and $0$-part (where each of the
intersections may be empty), and interpreting this as a mixture of possibly
three distinct distributions, we can also allow general $\R$-valued distributions
as factors in multiplications; the result can then possibly be a mixture of a
Dirac distribution in $0$ and an absolutely continuous distribution.
For division we note that for distributions with positive support,
{\tt X/Y=exp(log(X)-log(Y))}, and similar arguments also allow us to cover
powers, i.e., expressions like \code{X}$\hat{\hphantom i}$\code{Y}. As an example, let us see
how the distribution of $X=N\times P$ looks like if $N\sim{\cal N}(0,1)$ and
$P\sim{\rm Poisson}(\lambda)$:
\begin{Schunk}
\begin{Sinput}
R> X <- Norm() * Pois(lambda = 1)
R> q(X)(.25)
\end{Sinput}
\begin{Soutput}
[1] -0.3471003
\end{Soutput}
\begin{Sinput}
R> p(X)(1:3)
\end{Sinput}
\begin{Soutput}
[1] 0.8545304 0.9409595 0.9729868
\end{Soutput}
\begin{Sinput}
R> r(X)(5)
\end{Sinput}
\begin{Soutput}
[1] 0.1811465 0.0000000 0.7561025 0.0000000 0.4428234
\end{Soutput}
\begin{Sinput}
R> plot(X, cex.inner = 1, to.draw.arg = c(1,2),
+       inner = c("cdf", "quantile function"))
\end{Sinput}
\end{Schunk}
\includegraphics{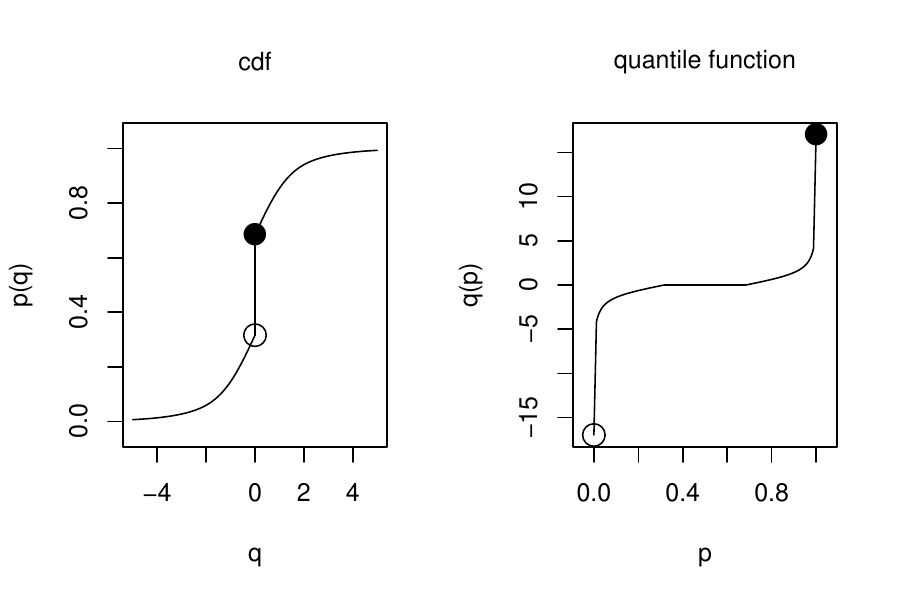}
%

\section{General purpose FFT algorithm}\label{sect4}
%
The main idea of our algorithm is to use DFT, which may be calculated
very fast by FFT. Hence, we start with a brief introduction to DFT and its
convolution property (cf.\ Theorem~\ref{ap.fft.dft.thm}) where we
follow Lesson~8 of \citet{Ga:Wi:99}. Afterwards, we describe the convolution
of cdf's/densities in Section~\ref{ap.fft.algo}.
\subsection{Discrete Fourier transformation (DFT)}\label{ap.fft.dft}
Let $m\in\N$ and let $(x_n)_{n\in\sZ}$ be a sequence of complex numbers with period $m$;
i.e., $x_{n+m}=x_n$ for all $n\in\Z$. Then, the DFT of order~$m$ is,
\begingroup \mathsurround0em\arraycolsep0em
\begin{eqnarray}
 &{\rm DFT}_m\colon\C^m\to\C^m,\;
 (x_0,x_1,\ldots,x_{m-1})\mapsto(\hat x_0,\hat x_1,\ldots,\hat x_{m-1})&\\
  \noalign{\noindent where \nopagebreak}\label{ap.fft.dft.dft}
  &\hat x_n {}={} \displaystyle\frac{1}{m}\sum_{j=0}^{m-1}x_j\omega_m^{jn}
  \qquad \omega_m {}={} e^{-2\pi i/m},\;i=\sqrt{-1}\,&
\end{eqnarray}\endgroup
We obtain the DFT $(\hat{x}_n)_{n\in\sZ}$ of $(x_n)_{n\in\sZ}$
by the periodic extension $\hat{x}_{n+m}=\hat{x}_n$ for all $n\in\Z$.
${\rm DFT}_m$ is represented by a matrix $\Omega_m$ with
entries $\omega_m^{jk}$ ($j,k=0,1,\ldots,m-1$) and inverse
$\Omega_m^{-1}=1/m\,\overline{\Omega}_m$ ($\overline{\Omega}_m$ the
conjugate ${\rm DFT}_m$); i.e., ${\rm DFT}_m$ is linear and bijective.
\begin{Rem}\rm
\par{\bf (a)} Computing $\hat{x}_0, \hat{x}_1,\ldots,\hat{x}_{m-1}$ directly
  from Equation~\ref{ap.fft.dft.dft}, requires $(m-1)^2$ complex
  multiplications and $m(m-1)$ complex additions. But,
  FFT as introduced by \citet{Co:Tu:65}, is of just order $m\log m$. It works
  best for the case $m=2^p$ ($p\in\N$); see Lesson~9 of \citet{Ga:Wi:99}.
  In case $m=2^{10}$, direct computation needs $1046529$ multiplications
  and $1047552$ additions, whereas FFT only requires  $4097$ multiplications
  and $10240$ additions; see also Table~9.1 (ibid.).
\par{\bf (b)} If $(x_n)_{n\in\Z}$ is a sequence of real numbers, it is
  possible to reduce the cost of computation by half; cf.\  Section~8.3
  of \citet{Ga:Wi:99}.
\par{\bf (c)} FFT is available in \proglang{R} as function \code{fft}.
\end{Rem}
For DFTs we have the following convolution theorem:
\begin{Thm}\sl\label{ap.fft.dft.thm}
Let $x=(x_n)_{n\in\sZ}$ and $y=(y_n)_{n\in\sZ}$ be two sequences
of complex numbers with period $m$ and let $\hat{x}=(\hat{x}_n)_{n\in\sZ}$
and $\hat{y}=(\hat{y}_n)_{n\in\sZ}$ be the corresponding DFTs.
Then, the circular convolution product of $x$ and $y$ is defined as,
\begin{equation}\label{ap.fft.conv.form}
  x * y = \bigg(\sum_{j=0}^{m-1}x_j y_{n-j}\bigg)_{n\in\Z}
\end{equation}
and it holds,
\begin{equation}
  \hat{z} = m\,\hat{x}\,\hat{y} \qquad  \mbox{with} \qquad z=x*y
\end{equation}
where $\hat{x}\,\hat{y} = (\hat{x}_n\hat{y}_n)_{n\in\sZ}$.
\end{Thm}
The proof
is standard; see for instance \citet[Theorem~C.1.2]{Koh:04di}.
This Theorem implies the following result for $N$-fold convolution
products.
\begin{Prop}\sl\label{ap.fft.dft.prop}
Let $x=(x_n)_{n\in\sZ}$ be a sequence of complex numbers with period~$m$
and let $\hat{x}=(\hat{x}_n)_{n\in\sZ}$ be the corresponding DFT.
Then, it holds,
\begin{equation}
  \widehat{*_{i=1}^N x} = m^{N-1}\,\hat{x}^N\qquad N\in\N
\end{equation}
\end{Prop}
 The proof immediately follows from Theorem~\ref{ap.fft.dft.thm} by induction.
\subsection{Convolution algorithm}\label{ap.fft.algo}
DFT is formulated for discrete (equidistant) sequences of complex numbers,
 as which we may interpret the probability function of the following
special integer lattice distributions
\begingroup \mathsurround0em\arraycolsep0em
\begin{eqnarray}
  F_i(x) &{}={}& \sum_{j=0}^{m-1}p_{i,j}\Jc_{[j,\infty)}(x) \qquad i=1,2\\
  \noalign{\noindent with \nopagebreak}
  p_{i,j} &{}\ge{}& 0 \qquad j=0,1,\ldots,m-1
  \qquad \textstyle\sum\limits_{j=0}^{m-1}p_{i,j}=1
\end{eqnarray}\endgroup
where $x\in\R$ and $m=2^q$ ($q\in\N$). We extend
$p_{i,j}$ ($i=1,2$, $j=0,\ldots,m-1$) to two sequences
$p_i=(p_{i,n})_{n\in\sZ}$ of real numbers with period $2m$ via,
\begingroup \mathsurround0em\arraycolsep0em
\begin{eqnarray}
  p_{i,j} &{}={}& 0 \qquad i=1,2 \qquad j=m, \ldots, 2m-1
  \qquad\mbox{(zero padding)}\\
  \noalign{\noindent and \nopagebreak}
  p_{i,k+2m} &{}={}& p_{i,k} \qquad \forall\,k\in\Z
\end{eqnarray}\endgroup
Then, the convolution $F$ of $F_1$ and $F_2$ is an integer lattice
distribution given by
\begingroup \mathsurround0em\arraycolsep0em
\begin{eqnarray}
  F(x) {}={} (F_1*F_2)(x)
  &{}={}&
  \label{ap.fft.algo.conv}
  \sum_{j=0}^{2m-1}\pi_j\Jc_{[j,\infty)}(x)
  \qquad\mbox{with}\qquad
  \pi_j {}:={} \sum_{k=0}^{2m-1} p_{1,k}p_{2,j-k}\quad\qquad
\end{eqnarray}\endgroup
where in particular $\pi_{2m-1}=0$. Hence, in view of Theorem~\ref{ap.fft.dft.thm},
$\pi=(\pi_n)_{n\in\sZ} = p_1 * p_2$ and we can compute $\pi$ using FFT
and its inverse. This result forms the basis of Algorithm~\ref{ap.fft.algo.algo}.\\


As it stands, Algorithm~\ref{ap.fft.algo.algo} will be presented for the
case of absolutely continuous distributions, but with slight and obvious
modifications this algorithm works for quite general distributions;
for more details see also Section~\ref{ap.fft.algo.rem}.
\begin{algo}\sf\label{ap.fft.algo.algo}
Assume two absolutely continuous distributions $F_1, F_2$ on $\R$.
\begin{description}
\item[Step 1: (Truncation)]{\hspace{1em}}\\
If the support of $F_i$ ($i=1,2$) is unbounded or ``too large'',
we define numbers $A_i, B_i\in\R$, for given $\varepsilon > 0$, such that,
\begin{equation}
  F_i\big((-\infty,A_i)\big)=\frac{\varepsilon}{2} \qquad \mbox{and}\qquad
  F_i\big((B_i,\infty)\big)=\frac{\varepsilon}{2}
\end{equation}
and set $A=\min\{A_1,A_2\}$ and $B=\max\{B_1,B_2\}$. If this is not the case,
we define $A:=\min\{F_1^{-1}(0),F_2^{-1}(0)\}$ and $B:=\max\{F_1^{-1}(1),F_2^{-1}(1)\}$
where $F_i^{-1}$ ($i=1,2$) are the quantile functions of $F_i$.
\item[Step 2: (Discretization on a real grid)]{\hspace{1em}}\\
Given $m=2^q$ ($q\in\N$) and $F_i$ ($i=1,2$), we define
the lattice distributions\index{distribution!lattice}
\begingroup \mathsurround0em\arraycolsep0em
\begin{eqnarray}
  G_i(x) &{}:={}& \displaystyle\sum_{j=0}^{m-1}p_{i,j}\Jc_{[A+(j+0.5)h,\infty)}(x)
  \qquad h {}={} \frac{B-A}{m}\\
  \noalign{$\qquad\:$ where \nopagebreak}
  p_{i,j} &{}={}& F_i\big([A+jh,A+(j+1)h]\big)
\end{eqnarray}\endgroup
for $j=0,1,\ldots,m-1$.
\item[Step 3: (Transformation to an integer grid)]{\hspace{1em}}\\
Based on $G_i$ ($i=1,2$), we define the integer lattice
distributions\index{distribution!integer lattice}
\begin{equation}
  \tilde G_i(x) := \sum_{j=0}^{m-1}p_{i,j}\Jc_{[j,\infty)}(x)\qquad i=1,2
\end{equation}
and extend $p_{i,j}$ ($i=1,2$, $j=0,\ldots,m-1$) to two sequences
$p_i=(p_{i,n})_{n\in\sZ}$ of real numbers with period $2m$ via,
\begingroup \mathsurround0em\arraycolsep0em
\begin{eqnarray}
  &p_{i,j} {}={} 0 \qquad i=1,2 \qquad j=m, \ldots, 2m-1
  \qquad\mbox{(zero padding)}&\\
  \noalign{$\qquad\:$ and \nopagebreak}
  &p_{i,k+2m} {}={} p_{i,k} \qquad \forall\,k\in\Z&
\end{eqnarray}\endgroup
\item[Step 4: (Convolution by FFT on integer grid)]{\hspace{1em}}\\
We calculate $\tilde G = \tilde G_1*\tilde G_2$
by FFT and its inverse\index{FFT!inverse} as given
in~Equation~\ref{ap.fft.algo.conv}; i.e.,
\begin{equation}
  \tilde G(x) = \sum_{j=0}^{2m-1}\pi_j\Jc_{[j,\infty)}(x)
  \qquad
  \pi_j := \sum_{k=0}^{2m-1} p_{1,k}p_{2,j-k}
\end{equation}
where in particular $\pi_{2m-1}=0$.
\item[Step 5: (Back-transformation to real grid)]{\hspace{1em}}\\
Given $\tilde G$, we obtain $G=G_1*G_2$ by,
\begin{equation}
  G(x) = \sum_{j=0}^{2m-2}\pi_j\Jc_{[2A + (j+1.5)h,\infty)}(x)
\end{equation}
That is, we additionally use a continuity correction of $h/2$,
which improves the accuracy of the results.
\item[Step 6: (Smoothing)]{\hspace{1em}}\\
Next, we use interpolation of the values of $G$ on
$\{2A, 2A+1.5 h,\ldots,2B-0.5h, 2B\}$ by linear functions to get a
continuous approximation $F^\natural$ of $F=F_1*F_2$.
We obtain a continuous approximation $f^\natural$ of the density $f$
of $F$ by multiplying $\{0,\pi_0,\pi_1,\ldots,\pi_{2m-2},0\}$ by~$h$ and
interpolating these values on the grid $\{2A,2A+h,\ldots,2B-h,2B\}$
(no continuity correction) using linear functions.
\item[Step 7: (Standardization)]{\hspace{1em}}\\
To make sure that the approximation $F^\natural$ is indeed a probability
distribution, we standardize~$F^\natural$ and~$f^\natural$ by
$F^\natural\big([2A, 2B]\big)$ and $\int f^\natural(x)\,dx$,
respectively, where $\int f^\natural(x)\,dx$ may be calculated numerically
exactly, since $f^\natural$ is a piecewise linear function.
\end{description}
\end{algo}
For some instructive examples like the computation of (an approximation to)
the stationary regressor distribution of an AR(1) process, together with
corresponding \proglang{R} sources see \citet{RMandistr}.
\subsection{Ramifications and extensions of this algorithm}\label{ap.fft.algo.rem}
{\bf Algorithm~\ref{ap.fft.algo.algo} for lattice distributions: }
Obviously, Algorithm~\ref{ap.fft.algo.algo} applies to lattice
distributions\index{distribution!lattice} $F_1, F_2$
 on~$\R$ defined on the same grid. In this case the algorithm
 essentially reduces to steps 1-5 and 7. Moreover, the results are
  numerically exact if the lattice distributions have finite support;
  cf.\  Section~\ref{ap.fft.check}. In this case the algorithm consists
  only of steps 2-5.\smallskip\\
{\bf Specification of ``too large'': }In step 1, a support is considered as ``too large'' if
  a uniform grid with a reasonable step-length produces too many grid
  points. In the same sense, the loss of mass included in step~1 of
  Algorithm~\ref{ap.fft.algo.algo} is, to some extent, controllable and
  in many cases negligible.\smallskip\\
{\bf Richardson Extrapolation: }A technique to enhance the accuracy of
  Algorithm~\ref{ap.fft.algo.algo} for given $q$ is extrapolation. But,
  for this to work properly, we need additional smoothness conditions
  for the densities. We could take this into account by introducing
  a new subclass \code{SmoothDistribution} for distributions with
  sufficiently smooth densities and a corresponding new method for
  the operator ``\code{+}''; see also Section~\ref{ap.fft.algo.rem0}.
  \smallskip\\
{\bf Exponential Tilting: }As a wrap-around effect, summation modulo~$m$
  (cf.\ Equation~\ref{ap.fft.conv.form}) induces an aliasing error.
  Especially for heavy-tailed distributions -- again at the cost of
  additional smoothness conditions for the densities --
  Algorithm~\ref{ap.fft.algo.algo} can thus be improved
  by a suitable change of measure (exponential tilting). So one might conceive a
  further subclass \code{HeavyTailedSmoothDistribution} and overload ``\code{+}''
  for objects of these classes using exponential tilting; see also
  Section~\ref{ap.fft.algo.rem0}.\smallskip\\
{\bf Modification for M-Estimators: }In view of Proposition~\ref{ap.fft.dft.prop},
  Algorithm~\ref{ap.fft.algo.algo} may easily be modified
  to compute an approximation of the exact finite-sample distribution
  of M estimates, compare \citet{Ru:Ko:04}. In the cited
  reference, we compare the results obtainable with this modified algorithm
  to other approximations of the exact finite-sample distribution
  of M estimates, like the saddle point approximation
  and higher order asymptotics.

\section{Connections to other approaches}\label{otherapp} 
\subsection{Algorithms based on DFT}\label{ap.fft.algo.rem0}
A very similar algorithm was proposed by~\citet{Bert:81}
to numerically evaluate compound distributions in insurance mathematics
where he assumes claim size distributions of lattice type.
Numerical examples and comparisons to other methods can be found
in~\citet{Bue:84} and~\citet{FB:87}.

A mathematical formulation of the corresponding algorithm is included in
\citet{GH:99}. However, the main purpose of their article is the investigation
of the aliasing error. In case of a claim size distribution of lattice type
they obtain a simple general bound for this error and show that it can
be eliminated by exponential tilting. But, even without the smoothness
assumptions needed for exponential tilting, the aliasing error can
also be made very small if we choose $\varepsilon$ in step~1 of
Algorithm~\ref{ap.fft.algo.algo} small enough and $q$ in step~2 large
enough. Thus, in many cases this effect is negligible.

Moreover, if one considers absolutely continuous probability
distributions, an initial discretization step is necessary; see
Step~2 of Algorithm~\ref{ap.fft.algo.algo}. The corresponding error
is studied in \citet{GH:00} and it is shown that this error, under certain
smoothness conditions, can be reduced by an extrapolation technique
(Richardson extrapolation).

Efficient and precise algorithms based on FFT for the convolution of
heavy-tailed distributions are considered in \citet{Sch:Tem:08}.

In \citet{EGP:93} the authors describe how one can use FFT
to determine various quantities of interest in risk theory and insurance
mathematics including the computation of the total claim size distribution,
the mean and the variance of the process and the probability of ruin.
Moreover, using FFT it is also possible to find the stationary waiting
time distribution for a given customer inter-arrival time distribution
and a given service time distribution in the G/G/1 queueing model;
see \citet{G:91}.
%
\subsection{Other algorithms}
For continuous distributions, instead of starting with a discretization
of the cdf right away, we could also use the actual characteristic
functions, i.e., the Fourier transformations of the corresponding distributions
which then get inverted by the usual Fourier inversion formulae, see
e.g., \citet[Sec.6.2]{Ch:74}. As coined
by Th.~Lumely in a posting to
\href{http://tolstoy.newcastle.edu.au/R/e2/help/07/03/13491.html}{\tt r-help}
on March 29, 2007, this is in particular useful if there are closed form
expressions for the characteristic functions as for instance for
linear combination of independent $\chi^2$-distributions.

On the other hand, inverting characteristic functions is not a cure-all procedure
either, as may be seen when considering convolution powers of the uniform distribution
on $[-1/2,1/2]$: The corresponding characteristic functions are $(\sin(t/2)/t)^n$
which does if na\"ively inverted cause quite some numerical problems.
A more comprehensive account of this approach can be found in
\citet{Cav:78}, \citet{Ab:Wh:92} and \citet{Abate:95}.

Similarily, but with a restricted application range due to integrability one
could stay on the real line using Laplace transformations; see for instance
\citet{Ab:Wh:92} and \citet{Abate:95}.

In actuarial science, recursive schemes to compute convolution powers,
the so-called \textit{Panjer recursions}, have been in use for a long time.
As \cite{TW08} show, these recursive methods are slower than FFT when a
sufficient precision of the estimated quantile is needed.
\section{Accuracy and computational efficiency of our algorithm}\label{ap.fft.check}
To assess the accuracy and computational efficiency of our algorithm,
we present checks for $n$-fold  convolution products where the exact results
are known. In addition, we approximate probabilities of non-central
$\chi^2$-distributions.
\subsection{Accuracy}
We determine
the precision of the convolution algorithm in terms of the total variation
distance of the densities,
\begin{equation}
  d_v(P,Q) = \Tfrac{1}{2}\int|p - q|\,d\mu = \sup\limits_{B\in\B}\big|P(B)-Q(B)\big|
\end{equation}
where $P,Q\in{\cal M}_1(\B)$ with $dP=p\,d\mu$, $dQ=q\,d\mu$ for some
$\sigma$-finite measure $\mu$ on $(\R,\B)$
and the Kolmogorov distance of the cumulative distribution functions,
\begin{equation}
  d_\kappa(P,Q) =
  \sup\limits_{t\in\R}\big|P\big((-\infty,t]\big) - Q\big((-\infty,t]\big)\big|
\end{equation}
Obviously, $d_\kappa\le d_v$ as the supremum in case of the total variation
distance is taken over more sets. In the sequel $d_v^\natural$ and
$d_\kappa^\natural$ stand for the numerical approximations of $d_v$
and $d_\kappa$. Due to numerical inaccuracies we obtain
$d_\kappa^\natural > d_v^\natural$ in some cases.
\par The first example treats Binomial distributions and shows that the
convolution algorithm is very accurate for integer lattice
distributions with finite support.
\begin{Exa}\rm\label{ap.fft.check.bin}
Assume $F={\rm Bin}\,(k,p)$ with $k\in\N$ and $p\in(0,1)$. Then, the
$n$-fold convolution product is $F^{*n}={\rm Bin}\,(nk, p)$ ($n\in\N$).
Let $f_n$ and $f^\natural$ be the probability functions of $F^{*n}$
and $F^\natural$, respectively. Then, we may determine $d_v^\natural$
and $d_\kappa^\natural$ numerically exact by,
\begingroup \mathsurround0em\arraycolsep0em
\begin{eqnarray}
  d_v^\natural(F,F^\natural) &{}={}&
  \frac{1}{2}\sum_{j=0}^{nk}|f_n(j)-f^\natural(j)|\\
  \noalign{\noindent and \nopagebreak}
  d_\kappa^\natural(F,F^\natural) &{}={}& \max_{j\in\{0,\ldots,nk\}}
  \big|F^{*n}([0,j])-F^\natural([0,j])\big|
\end{eqnarray}\endgroup
We obtain the results contained in Table~\ref{ap.fft.check.tab.bin} which
show that Algorithm~\ref{ap.fft.algo.algo} is very accurate in case
of binomial distributions, where the values of $k$ and $p$ are chosen
arbitrarily. To get the corresponding results we use our \proglang{R}
packages \pkg{distr} and \pkg{distrEx}. For example
\begin{Schunk}
\begin{Sinput}
R> library("distrEx")
R> distroptions(TruncQuantile = 1e-15)
R> B1 <- Binom(size = 30, prob = 0.8)
R> B2 <- convpow(B1, N = 10)
R> D1 <- as(B1, "LatticeDistribution")
R> D2 <- convpow(D1, N = 10)
R> TotalVarDist(B2, D2)
\end{Sinput}
\begin{Soutput}
total variation distance
            2.273135e-15
\end{Soutput}
\begin{Sinput}
R> KolmogorovDist(B2, D2)
\end{Sinput}
\begin{Soutput}
Kolmogorov distance
       1.249001e-15
\end{Soutput}
\end{Schunk}
where \code{B2} is computed using the exact formula and \code{D2} is
the approximation via FFT. To increase accuracy we change the default
value of option \code{TruncQuantile} from $1{\rm e}{-5}$ to $1{\rm e}{-15}$.
\begin{table}[!ht]
\begin{center}
\begin{tabular}[c]{r||r|c||c|c}
 $n$ & $k$ & $p$ &    $d_v^\natural$ & $d_\kappa^\natural$ \\\hline\hline
   2 &  10 & 0.5 & $3.3{\rm e}{-16}$ & $2.2{\rm e}{-16}$  \\ \hline
   5 &  20 & 0.7 & $1.7{\rm e}{-15}$ & $9.6{\rm e}{-16}$  \\ \hline
  10 &  30 & 0.8 & $2.6{\rm e}{-15}$ & $1.1{\rm e}{-15}$  \\ \hline
 100 &  15 & 0.2 & $5.3{\rm e}{-15}$ & $4.3{\rm e}{-15}$  \\ \hline
1000 &  50 & 0.4 & $8.3{\rm e}{-13}$ & $4.2{\rm e}{-13}$ \\ \hline
\end{tabular}
\caption[Precision of the Convolution of Binomial Distributions via FFT]%
{\label{ap.fft.check.tab.bin} Precision of the convolution of binomial distributions via FFT;
see Example~\ref{ap.fft.check.bin}.}
\end{center}
\end{table}
\end{Exa}
In case of the Poisson distribution the results of the convolution algorithm
turn out to be very accurate, too.
\begin{Exa}\rm\label{ap.fft.check.pois}
We consider $F={\rm Pois}\,(\lambda)$ with $\lambda\in(0,\infty)$ where
$F^{*n}={\rm Pois}\,(n\lambda)$ ($n\in\N$). Since the support of $F$
is $\N_0$, we use $A=0$ and $B=F^{-1}(1-1{\rm e}{-15})$ in step~1 of
Algorithm~\ref{ap.fft.algo.algo} and determine $d_v^\natural$ and
$d_\kappa^\natural$ numerically exact by,
\begingroup \mathsurround0em\arraycolsep0em
\begin{eqnarray}
  d_v^\natural(F,F^\natural) &{}={}& \frac{1}{2}\sum_{j=0}^{M}|f_n(j)-f^\natural(j)|\\
  \noalign{\noindent and \nopagebreak}
  d_\kappa^\natural(F,F^\natural) &{}={}& \max_{j\in\{0,\ldots,M\}}
  \big|F^{*n}([0,j])-F^\natural([0,j])\big|
\end{eqnarray}\endgroup
where $M$ is the $1-1{\rm e}{-15}$ quantile of~$F^{*n}$.
We obtain the results contained in Table~\ref{ap.fft.check.tab.pois} which
demonstrate the high precision of the convolution algorithm in case of
Poisson distributions where the parameter $\lambda$ is chosen arbitrarily.
The results can be obtained via our \proglang{R} packages \pkg{distr} and
\pkg{distrEx} analogously to the Binomial case.
\begin{Schunk}
\begin{Sinput}
R> library("distrEx")
R> distroptions(TruncQuantile = 1e-15)
R> P1 <- Pois(lambda = 15)
R> P2 <- convpow(P1, N = 100)
R> D1 <- as(P1, "LatticeDistribution")
R> D2 <- convpow(D1, N = 100)
R> TotalVarDist(P2, D2)
\end{Sinput}
\begin{Soutput}
total variation distance
            1.616895e-13
\end{Soutput}
\begin{Sinput}
R> KolmogorovDist(P2, D2)
\end{Sinput}
\begin{Soutput}
Kolmogorov distance
        8.85958e-14
\end{Soutput}
\end{Schunk}
\begin{table}[!ht]
\begin{center}
\begin{tabular}{r||r||c|c}
 $n$ &  $\lambda$ &      $d_v^\natural$ & $d_\kappa^\natural$ \\ \hline\hline
   2 &        0.1 & $2.9{\rm e}{-16}$ & $2.2{\rm e}{-16}$ \\ \hline
   5 &       10.0 & $3.7{\rm e}{-15}$ & $3.1{\rm e}{-15}$ \\ \hline
  10 &        7.5 & $4.0{\rm e}{-15}$ & $4.0{\rm e}{-15}$ \\ \hline
 100 &       15.0 & $1.8{\rm e}{-13}$ & $1.0{\rm e}{-13}$ \\ \hline
1000 &       50.0 & $2.0{\rm e}{-11}$ & $1.0{\rm e}{-11}$ \\ \hline
\end{tabular}
\caption[Precision of the Convolution of Poisson Distributions via FFT]%
{\label{ap.fft.check.tab.pois}%
Precision of the convolution of Poisson distributions via FFT; see
Example~\ref{ap.fft.check.pois}.}
\end{center}
\end{table}
\end{Exa}
In the next two examples we consider the convolution of absolutely continuous
distributions. We determine the total variation distance
$d_v^\natural(F,F^\natural)$ by numerical integration using the \proglang{R}
function \code{integrate}. To compute an approximation of the Kolmogorov
distance, we evaluate $d_\kappa^\natural(F,F^\natural)$ on a grid obtained
by the union of a deterministic grid of size $1{\rm e}{05}$ and two random
grids consisting of $1{\rm e}{05}$ pseudo-random numbers of the considered
distributions. We first present the results for normal distributions.
\begin{Exa}\rm\label{ap.fft.check.norm}
Assume $F={\cal N}\,(\mu,\sigma^2)$ with $\mu\in\R$ and $\sigma\in(0,\infty)$.
Then it holds, $F^{*n}={\cal N}\,(n\mu, n\sigma^2)$
($n\in\N$). Starting with ${\cal N}\,(0,1)$ and $A$ and $B$ as defined in
step~1 of Algorithm~\ref{ap.fft.algo.algo} we obtain $\tilde A = \sigma A + \mu$
and $\tilde B = \sigma B + \mu$ in case of ${\cal N}\,(\mu, \sigma^2)$.
That is, the grid transforms the same way as the normal distributions do.
Thus, we expect the precision of the results to be independent of $\mu$
and $\sigma$. This is indeed confirmed by the numerical calculations; see
Table~\ref{ap.fft.check.tab.norm1}. We therefore may consider $\mu=0$ and
$\sigma=1$ for the study of the accuracy of the convolution algorithm
subject to $n\in\N$, $\varepsilon > 0$ (step~1) and $q\in\N$ (step~2).
The results included in Table~\ref{ap.fft.check.tab.norm2} show that the
precision is almost independent of $n$. It mainly depends on $q$ where the
maximum accuracy, we can reach, is of order $\varepsilon$. The results can be
computed with our \proglang{R} packages \pkg{distr} and \pkg{distrEx} similarily
to the Binomial and Poisson case.
\begin{Schunk}
\begin{Sinput}
R> library("distrEx")
R> distroptions(TruncQuantile = 1e-10)
R> distroptions(DefaultNrFFTGridPointsExponent = 14)
R> N1 <- Norm(mean = 0, sd = 1)
R> N2 <- convpow(N1, N = 2)
R> D1 <- as(N1, "AbscontDistribution")
R> D2 <- convpow(D1, N = 2)
R> distroptions(TruncQuantile = 1e-15)
R> TotalVarDist(N2, D2, rel.tol = 1e-10)
\end{Sinput}
\begin{Soutput}
total variation distance
            9.806121e-08
\end{Soutput}
\begin{Sinput}
R> KolmogorovDist(N2, D2)
\end{Sinput}
\begin{Soutput}
Kolmogorov distance
       1.700898e-07
\end{Soutput}
\end{Schunk}
\begin{table}[!ht]
\begin{center}
\begin{tabular}{r||c|r||r|r||c|c}
  $n$ & $\varepsilon$    & $q$  & $\mu$  & $\sigma$ &    $d_v^\natural$   & $d_\kappa^\natural$ \\ \hline\hline
      &                  &      &  -10.0 &    100.0 & $1.2{\rm e}{-06}$   & $2.1{\rm e}{-06}$ \\ \cline{4-7}
      &                  &      &   -2.0 &      5.0 & $1.2{\rm e}{-06}$   & $2.1{\rm e}{-06}$ \\ \cline{4-7}
  $2$ & $1{\rm e}{-08}$  & $12$ &    0.0 &      1.0 & $1.2{\rm e}{-06}$   & $2.1{\rm e}{-06}$ \\ \cline{4-7}
      &                  &      &    1.0 &     50.0 & $1.2{\rm e}{-06}$   & $2.1{\rm e}{-06}$ \\ \cline{4-7}
      &                  &      &  100.0 &   1000.0 & $1.2{\rm e}{-06}$   & $2.1{\rm e}{-06}$ \\ \hline
\end{tabular}
\caption[Precision of the Convolution of Normal Distributions via FFT
is Independent of the Parameters $\mu$ and $\sigma$]%
{\label{ap.fft.check.tab.norm1}%
Precision of the convolution of normal distributions via
FFT is independent of the parameters $\mu$ and $\sigma$;
see Example~\ref{ap.fft.check.norm}.}
\end{center}
\end{table}
\begin{table}[!ht]
\begin{center}
\begin{tabular}{r||c|r||c|c}
  $n$ & $\varepsilon$    & $q$ &     $d_v^\natural$ & $d_\kappa^\natural$          \\ \hline\hline
      &                  &   8 & $2.2{\rm e}{-04}$  & $3.9{\rm e}{-04}$  \\ \cline{3-5}
      & $1{\rm e}{-06}$  &  10 & $1.3{\rm e}{-05}$  & $2.3{\rm e}{-05}$  \\ \cline{3-5}
      &                  &  12 & $3.5{\rm e}{-06}$  & $1.8{\rm e}{-06}$  \\ \cline{2-5}
      &                  &  10 & $1.9{\rm e}{-05}$  & $3.4{\rm e}{-05}$  \\ \cline{3-5}
   2  & $1{\rm e}{-08}$  &  12 & $1.2{\rm e}{-06}$  & $2.1{\rm e}{-06}$  \\ \cline{3-5}
      &                  &  14 & $8.5{\rm e}{-08}$  & $1.2{\rm e}{-07}$  \\ \cline{2-5}
      &                  &  12 & $1.6{\rm e}{-06}$  & $2.7{\rm e}{-06}$  \\ \cline{3-5}
      & $1{\rm e}{-10}$  &  14 & $9.8{\rm e}{-08}$  & $1.7{\rm e}{-07}$  \\ \cline{3-5}
      &                  &  18 & $5.2{\rm e}{-10}$  & $5.3{\rm e}{-10}$ \\ \hline
   5  & $1{\rm e}{-08}$  &  12 & $3.4{\rm e}{-06}$  & $9.7{\rm e}{-04}$  \\ \cline{3-5}
      &                  &  16 & $6.6{\rm e}{-08}$  & $6.1{\rm e}{-05}$  \\ \hline
  10  & $1{\rm e}{-08}$  &  12 & $1.1{\rm e}{-05}$  & $1.1{\rm e}{-05}$  \\ \cline{3-5}
      &                  &  16 & $6.3{\rm e}{-08}$  & $3.5{\rm e}{-08}$  \\ \hline
  50  & $1{\rm e}{-08}$  &  12 & $1.6{\rm e}{-04}$  & $9.6{\rm e}{-05}$  \\ \cline{3-5}
      &                  &  18 & $1.0{\rm e}{-07}$  & $5.3{\rm e}{-08}$ \\ \hline
\end{tabular}
\caption[Precision of the Convolution of Normal Distributions via FFT]%
{\label{ap.fft.check.tab.norm2}%
Precision of the convolution of normal distributions via FFT; see
Example~\ref{ap.fft.check.norm}.}
\end{center}
\end{table}
\end{Exa}
Our last example treats the convolution of
exponential distributions\index{distribution!Exponential} which leads to
gamma distributions\index{distribution!Gamma}.
\begin{Exa}\rm\label{ap.fft.check.exp}
We consider $F={\rm Exp}\,(\lambda)=\Gamma(1,\lambda)$ with
$\lambda\in(0,\infty)$. Then it holds, $F^{*n}=\Gamma\,(n, \lambda)$ ($n\in\N$).
Analogously to the normal case (cf.\ Example~\ref{ap.fft.check.norm}), the
grid transforms the same as the exponential distributions do; i.e.,
$\tilde A = 1/\!\lambda\, A$ and $\tilde B = 1/\!\lambda\, B$. Thus, we
expect the precision of the results to be independent of $\lambda$.
This is again confirmed by our numerical computations; see
Table~\ref{ap.fft.check.tab.exp1}. Next we study the dependence of
the accuracy of Algorithm~\ref{ap.fft.algo.algo} on $n\in\N$, $\varepsilon > 0$
and $q\in\N$ where we may choose $\lambda=1.0$. As in
Example~\ref{ap.fft.check.norm} the precision is almost independent of $n$.
It mainly depends on $q$ where the maximum accuracy, we can reach,
is of order~$\varepsilon$; see Table~\ref{ap.fft.check.tab.exp2}. The results
can be computed with our \proglang{R} packages \pkg{distr} and \pkg{distrEx}
similarily to the previous cases.
\begin{Schunk}
\begin{Sinput}
R> library("distrEx")
R> distroptions(TruncQuantile = 1e-8)
R> distroptions(DefaultNrFFTGridPointsExponent = 16)
R> E1 <- Exp(rate = 1)
R> E2 <- convpow(E1, N = 5)
R> D1 <- as(E1, "AbscontDistribution")
R> D2 <- convpow(D1, N = 5)
R> distroptions(TruncQuantile = 1e-15)
R> TotalVarDist(E2, D2, rel.tol = 1e-10)
\end{Sinput}
\begin{Soutput}
total variation distance
             1.39883e-07
\end{Soutput}
\begin{Sinput}
R> KolmogorovDist(E2, D2)
\end{Sinput}
\begin{Soutput}
Kolmogorov distance
       9.455245e-08
\end{Soutput}
\end{Schunk}
\begin{table}[!ht]
\begin{center}
\begin{tabular}{r||c|r||r|c|c}
  $n$ & $\varepsilon$   & $q$ & $\lambda$ &    $d_v^\natural$   & $d_\kappa^\natural$ \\ \hline\hline
      &                 &     &      0.01 & $5.6{\rm e}{-06}$   & $4.0{\rm e}{-05}$  \\ \cline{4-6}
      &                 &     &       0.5 & $5.6{\rm e}{-06}$   & $4.0{\rm e}{-05}$  \\ \cline{4-6}
   2  & $1{\rm e}{-08}$ & 12  &       1.0 & $5.6{\rm e}{-06}$   & $4.0{\rm e}{-05}$  \\ \cline{4-6}
      &                 &     &       5.0 & $5.6{\rm e}{-06}$   & $4.0{\rm e}{-05}$  \\ \cline{4-6}
      &                 &     &      10.0 & $5.6{\rm e}{-06}$   & $4.0{\rm e}{-05}$ \\ \hline
\end{tabular}
\caption[Precision of the Convolution of Exponential Distributions via FFT
is Independent of the Parameter $\lambda$]%
{\label{ap.fft.check.tab.exp1}%
Precision of the convolution of exponential distributions via FFT is
independent of the parameter $\lambda$; see Example~\ref{ap.fft.check.exp}.}

\end{center}
\end{table}
\begin{table}[!ht]
\begin{center}
\begin{tabular}{r||c|r||c|c}
  $n$ & $\varepsilon$    & $q$ &    $d_v^\natural$  & $d_\kappa^\natural$ \\ \hline\hline
      &                  &   8 & $7.5{\rm e}{-04}$  & $4.7{\rm e}{-03}$  \\ \cline{3-5}
      & $1{\rm e}{-06}$  &  10 & $4.7{\rm e}{-05}$  & $3.4{\rm e}{-04}$  \\ \cline{3-5}
      &                  &  12 & $4.5{\rm e}{-06}$  & $2.2{\rm e}{-05}$  \\ \cline{2-5}
      &                  &  10 & $8.1{\rm e}{-05}$  & $6.0{\rm e}{-04}$  \\ \cline{3-5}
   2  & $1{\rm e}{-08}$  &  12 & $5.6{\rm e}{-06}$  & $4.0{\rm e}{-05}$  \\ \cline{3-5}
      &                  &  16 & $3.6{\rm e}{-08}$  & $1.6{\rm e}{-07}$  \\ \cline{2-5}
      &                  &  12 & $8.0{\rm e}{-06}$  & $6.2{\rm e}{-05}$  \\ \cline{3-5}
      & $1{\rm e}{-10}$  &  14 & $5.1{\rm e}{-07}$  & $3.9{\rm e}{-06}$  \\ \cline{3-5}
      &                  &  20 & $2.7{\rm e}{-10}$  & $9.6{\rm e}{-10}$ \\ \hline
   5  & $1{\rm e}{-08}$  &  12 & $2.7{\rm e}{-05}$  & $2.8{\rm e}{-05}$  \\ \cline{3-5}
      &                  &  16 & $1.4{\rm e}{-07}$  & $9.5{\rm e}{-08}$  \\ \hline
  10  & $1{\rm e}{-08}$  &  12 & $1.4{\rm e}{-04}$  & $1.4{\rm e}{-04}$  \\ \cline{3-5}
      &                  &  16 & $6.2{\rm e}{-07}$  & $5.3{\rm e}{-07}$  \\ \hline
  50  & $1{\rm e}{-08}$  &  12 & $4.9{\rm e}{-03}$  & $4.9{\rm e}{-03}$  \\ \cline{3-5}
      &                  &  20 & $3.8{\rm e}{-07}$  & $3.8{\rm e}{-07}$ \\ \hline
\end{tabular}
\caption[Precision of the Convolution of Exponential Distributions via FFT]%
{\label{ap.fft.check.tab.exp2}%
Precision of the convolution of exponential distributions via FFT; see
Example~\ref{ap.fft.check.exp}.}
\end{center}
\end{table}
\end{Exa}
\begin{Rem}\rm\label{ap.fft.check.rem}
  Example~\ref{ap.fft.check.exp} reveals one minor flaw
  of Algorithm~\ref{ap.fft.algo.algo}.
  The support~of $\Gamma(n,\lambda)$ is $[0,\infty)$ whereas the
  convolution algorithm is only very accurate in
  $[2A+(n/2+0.5)h,\ldots,2B-(n/2+0.5)h]$. That is, for small $n$ ($n\le 5$)
  the Kolmogorov distance\index{distance!Kolmogorov} is
  $F\big([0, 2A+(n/2+0.5)h)\big)-F^\natural\big([0, 2A+(n/2+0.5)h)\big)$.
  However, for bigger~$n$ this inaccuracy disappears as there is less
  and less mass in $[0,\,2A+(n/2+0.5)h)$. Moreover, since $(n/2+0.5)h$ is
  very small, this also causes the numerical inaccuracy of $d_v^\natural$
  for small $n$ and leads to $d_\kappa^\natural > d_v^\natural$.
\end{Rem}
\begin{Exa}\rm\label{ap.fft.check.chisq}
In this last example we show how our FFT approach can be used to compute
probabilities for non-central $\chi^2$-distributions where the exact
values are difficult to obtain.
Let $X$ be a non-central $\chi^2$ distributed random variable with {\tt df}
degrees of freedom and non-centrality parameter {\tt ncp}; i.e.,
$X\sim\chi^2_{\tt df}({\tt ncp})$. Our goal is to approximate the cdf
$P(X \le x)$ at $x\in(0,\infty)$. In Table~\ref{ap.fft.check.tab.chisq}
we give exact values of \citet{Pat:49} (Patnaik), approximations by
\citet{Itt:00} (Ittrich), approximations by function \texttt{pchisq} of
package \pkg{stats}, \citep{RMANUAL}, (R-Core) as well as the results of three FFT approaches
(FFT1--FFT3). In the first case (FFT1) we approximate $X$ by
\begin{equation}
 X \approx Z_1^2 + Z_2^2 + \ldots + Z_{\tt df}^2
 \qquad\mbox{with}\qquad Z_i\sim{\cal N}\,\left(\sqrt{\frac{{\tt ncp}}{{\tt df}}}, 1\right)
\end{equation}
Secondly (FFT2) we use
\begin{equation}
 X \approx Z_1^2 + Z_2^2 + \ldots + Z_{{\tt df}-1}^2 + Z_{\tt df}^2
\end{equation}
where $Z_i\sim{\cal N}\,(0,1)$ for $i=1,\ldots,{\tt df}-1$ and
$Z_{\tt df}\sim{\cal N}\,(\sqrt{{\tt ncp}}, 1)$. Our third approximation (FFT3)
reads
\begin{equation}
 X \approx Y + Z^2
\end{equation}
where $Y\sim\chi^2_{{\tt df}-1}(0)$ (a central $\chi^2$-distribution) and
$Z\sim{\cal N}\,(\sqrt{{\tt ncp}}, 1)$.\\
For the FFT computations we used $\varepsilon = 1{\rm e}{-08}$ and $q = 18$. All
three FFT approaches give very good approximations. In particular, FFT3 yields
results which have the same accuracy as \texttt{pchisq} and the approximation
of \citet{Itt:00}.
\begin{Schunk}
\begin{Sinput}
R> library("distr")
R> distroptions(withgaps = FALSE)
R> distroptions(TruncQuantile = 1e-8)
R> distroptions(DefaultNrFFTGridPointsExponent = 18)
R> df0 <- 4
R> ncp0 <- 4
R> x0 <- 1.765
R> Z <- Norm(mean = sqrt(ncp0/df0))
R> Z2 <- Z^2
R> res1 <- convpow(Z2, N = df0)
R> Z <- Norm()
R> Z2 <- Z^2
R> X2 <- convpow(Z2, N = df0-1)
R> Y2 <- Norm(mean = sqrt(ncp0))^2
R> res2 <- X2 + Y2
R> res3 <- Chisq(df = df0-1) + Y2
R> res <- c(p(res1)(x0), p(res2)(x0), p(res3)(x0),
+           pchisq(x0, df = df0, ncp = ncp0))
R> names(res) <- c("FFT1", "FFT2", "FFT3", "R")
R> res
\end{Sinput}
\begin{Soutput}
      FFT1       FFT2       FFT3          R
0.04999865 0.04999924 0.04999936 0.04999937
\end{Soutput}
\end{Schunk}
\begin{table}[!ht]
\begin{center}
\begin{small}
\begin{tabular}{r|r|r||c|c|c|c|c|c}
{\tt df} & {\tt ncp} & $x$      & Patnaik   & Ittrich     & R-Core           & FFT1        & FFT2        & FFT3        \\ \hline\hline
         &           & $1.765$  & $0.0500$  & $0.0499994$ & $.\,.\,\ldots\ldots .$ & $.\,.\,\ldots .\,.\, 87$ & $.\,.\,\ldots\ldots 2$  & $.\,.\,\ldots\ldots .$ \\ \cline{3-9}
         &  $4$      & $10.000$ & $0.7118$  & $0.7117928$ & $.\,.\,\ldots\ldots .$ & $.\,.\,\ldots\ldots 5$   & $.\,.\,\ldots\ldots .$ & $.\,.\,\ldots\ldots .$ \\ \cline{3-9}
    $4$  &           & $17.309$ & $0.9500$  & $0.9499957$ & $.\,.\,\ldots\ldots .$ & $.\,.\,\ldots\ldots .$ & $.\,.\,\ldots\ldots .$ & $.\,.\,\ldots\ldots .$ \\ \cline{3-9}
         &           & $24.000$ & $0.9925$  & $0.9924604$ & $.\,.\,\ldots\ldots .$ & $.\,.\,\ldots\ldots .$ & $.\,.\,\ldots\ldots .$ & $.\,.\,\ldots\ldots .$ \\ \cline{2-9}
         & $10$      & $10.000$ & $0.3148$  & $0.3148207$ & $.\,.\,\ldots\ldots .$ & $.\,.\,\ldots\ldots 4$ & $.\,.\,\ldots\ldots 6$ & $.\,.\,\ldots\ldots .$ \\ \hline
         &  $1$      & $4.000$  & $0.1628$  & $0.1628330$ & $.\,.\,\ldots\ldots .$ & $.\,.\,\ldots .\,.\, 13$ & $.\,.\,\ldots .\,.\, 15$ & $.\,.\,\ldots\ldots .$ \\ \cline{3-9}
         &           & $16.004$ & $0.9500$  & $0.9500015$ & $.\,.\,\ldots\ldots 6$ & $.\,.\,\ldots\ldots 4$ & $.\,.\,\ldots\ldots .$ & $.\,.\,\ldots\ldots 6$ \\ \cline{2-9}
    $7$  &           & $10.257$ & $0.0500$  & $0.0499942$ & $.\,.\,\ldots\ldots .$ & $.\,.\,\ldots .\,.\, 39$ & $.\,.\,\ldots .\,.\, 39$ & $.\,.\,\ldots\ldots .$ \\ \cline{3-9}
         & $16$      & $24.000$ & $0.5898$  & $0.5863368$ & $.\,.\,\ldots\ldots .$ & $.\,.\,\ldots\ldots 6$ & $.\,.\,\ldots\ldots 4$ & $.\,.\,\ldots\ldots .$ \\ \cline{3-9}
         &           & $38.970$ & $0.9500$  & $0.9499992$ & $.\,.\,\ldots\ldots .$ & $.\,.\,\ldots\ldots .$ & $.\,.\,\ldots\ldots .$ & $.\,.\,\ldots\ldots .$ \\ \hline
         &  $6$      & $24.000$ & $0.8187$  & $0.8173526$ & $.\,.\,\ldots\ldots .$ & $.\,.\,\ldots .\,.\, 10$ & $.\,.\,\ldots .\,.\, 11$ & $.\,.\,\ldots\ldots .$ \\ \cline{2-9}
    $12$ & $18$      & $24.000$ & $0.2901$  & $0.2900495$ & $.\,.\,\ldots\ldots .$ & $.\,.\,\ldots .\,.\, 73$ & $.\,.\,\ldots .\,.\, 71$ & $.\,.\,\ldots\ldots .$ \\ \hline
         &  $8$      & $30.000$ & $0.7880$  & $0.7880015$ & $.\,.\,\ldots\ldots .$ & $.\,.\;.\,.\,79948$ & $.\,.\;.\,.\,79994$ & $.\,.\,\ldots\ldots .$ \\ \cline{3-9}
    $16$ &           & $40.000$ & $0.9632$  & $0.9632255$ & $.\,.\,\ldots\ldots .$ & $.\,.\,\ldots .\,.\, 43$ & $.\,.\,\ldots\ldots 1$ & $.\,.\,\ldots\ldots .$ \\ \cline{2-9}
         & $32$      & $30.000$ & $0.0609$  & $0.0628420$ & $.\,.\,\ldots\ldots 1$ & $.\,.\,\ldots .\,392$ & $.\,.\,\ldots .\,.\, 09$ & $.\,.\,\ldots\ldots .$ \\ \cline{3-9}
         &           & $60.000$ & $0.8316$  & $0.8315635$ & $.\,.\,\ldots\ldots .$ & $.\,.\,\ldots .\,.\, 14$ & $.\,.\,\ldots .\,.\, 23$ & $.\,.\,\ldots\ldots .$ \\ \hline
         &           & $36.000$ & $0.1567$  & $0.1567111$ & $.\,.\,\ldots\ldots .$ & $.\,.\,\ldots .\,018$ & $.\,.\,\ldots .\,023$ & $.\,.\,\ldots\ldots .$ \\ \cline{3-9}
    $24$ & $24$      & $48.000$ & $0.5296$  & $0.5296284$ & $.\,.\,\ldots\ldots .$ & $.\,.\,\ldots .\,177$ & $.\,.\,\ldots .\,174$ & $.\,.\,\ldots\ldots .$ \\ \cline{3-9}
         &           & $72.000$ & $0.9667$  & $0.9666954$ & $.\,.\,\ldots\ldots .$ & $.\,.\,\ldots .\,.\, 44$ & $.\,.\,\ldots .\,.\, 41$ & $.\,.\,\ldots\ldots .$ \\ \hline
\end{tabular}
\end{small}
\caption[Approximations of the CDF of Non-central $\chi^2$-Distributions via FFT]%
{\label{ap.fft.check.tab.chisq}%
Approximations of the cdf of non-central $\chi^2$-distributions via FFT; see
Example~\ref{ap.fft.check.chisq}. [$\varepsilon = 1{\rm e}{-08}$, $q = 18$, only the decimal
places which are different to Ittrich are given]}
\end{center}
\end{table}
\end{Exa}
\subsection{Computational efficiency}
To judge the computational efficiency of our algorithm, let us
check it in a situation where the exact solution of the convolution
is known, i.e., at the $10$-fold convolution of independent $\chi^2_1(0)$
distributions. As timings are of course subject to hardware considerations
we report relative timings, where as reference we use the implementation
in \proglang{R} package \pkg{actuar}. As for general distributions,
\pkg{actuar} already needs probabilities evaluated on a grid, we have
to wrap the respective function \code{aggregateDist} into a function
\code{convActuar} first, providing a respective discretization.
\begin{Schunk}
\begin{Sinput}
R> gc()
R> library("actuar")
R> distroptions(TruncQuantile = 1e-5)
R> distroptions(DefaultNrFFTGridPointsExponent = 12)
R> convActuar <- function(N = 2, df = 1, ncp = 0,
+                         method = "lower"){
+      D1 <- Chisq(df = df, ncp = ncp)
+      lo <- getLow(D1)
+      up <- getUp(D1)
+      dGPExp <- getdistrOption("DefaultNrFFTGridPointsExponent")
+      m <- max(dGPExp - floor(log(N)/log(2)), 5)
+      M <- 2^m
+      h <- (up - lo)/M
+      probs <- discretize(pchisq(x, df = df, ncp = ncp),
+                          from = lo, to = up, by = h,
+                          method = method)
+      x <- seq(from = N*lo+N/2*h, to = N*up-N/2*h, by = h)
+      x <- c(x[1]-h, x[1], x+h)
+      dx <- aggregateDist(method = "convolution",
+                          model.freq = c(rep(0, N),1),
+                          model.sev = probs)
+      list(d = dx, x = x)
+  }
\end{Sinput}
\end{Schunk}
No matter which of the methods implemented in function \code{discretize}
of package \pkg{actuar}, i.e., \code{"rounding"}, \code{"lower"},
or \code{"upper"}, our algorithm compares fairly well as to both
timings and accuracy:

\begin{Schunk}
\begin{Sinput}
R> system.time(res1 <- convActuar(method = "rounding"))
\end{Sinput}
\begin{Soutput}
   user  system elapsed
      0       0       0
\end{Soutput}
\begin{Sinput}
R> D1 <- as(Chisq(), "AbscontDistribution")
R> system.time(D2 <- convpow(D1 = D1, N = 2))
\end{Sinput}
\begin{Soutput}
   user  system elapsed
   0.02    0.00    0.01
\end{Soutput}
\begin{Sinput}
R> summary(abs(res1$d(knots(res1$d)) - p(D2)(
+          res1$x[c(-1, -length(res1$x))])))
\end{Sinput}
\begin{Soutput}
     Min.   1st Qu.    Median      Mean   3rd Qu.      Max.
2.005e-05 2.005e-05 2.018e-05 1.905e-04 4.636e-05 2.910e-03
\end{Soutput}
\begin{Sinput}
R> system.time(res2 <- convActuar(method = "upper"))
\end{Sinput}
\begin{Soutput}
   user  system elapsed
      0       0       0
\end{Soutput}
\begin{Sinput}
R> summary(abs(res2$d(knots(res2$d)) - p(D2)(
+          res2$x[c(-1, -length(res2$x))])))
\end{Sinput}
\begin{Soutput}
     Min.   1st Qu.    Median      Mean   3rd Qu.      Max.
3.600e-08 1.968e-05 2.000e-05 7.301e-05 2.000e-05 1.317e-03
\end{Soutput}
\begin{Sinput}
R> system.time(res3 <- convActuar(method = "lower"))
\end{Sinput}
\begin{Soutput}
   user  system elapsed
   0.01    0.00    0.02
\end{Soutput}
\begin{Sinput}
R> summary(abs(res3$d(knots(res3$d)) - p(D2)(res3$x)))
\end{Sinput}
\begin{Soutput}
     Min.   1st Qu.    Median      Mean   3rd Qu.      Max.
0.000e+00 2.000e-05 2.017e-05 2.006e-04 4.664e-05 4.855e-03
\end{Soutput}
\end{Schunk}
To see the differences more clearly, let us repeat this 100 times.

\begin{Schunk}
\begin{Sinput}
R> speedref <- function(expr.ref, rep.times = 100){
+       ref.time <- system.time(for(i in 1:rep.times)
+                                 res <- eval(expr.ref))[1]
+       names(ref.time) <- NULL
+       return(list(res = res, ref.time = ref.time))
+  }
R> speedcheck <- function(expr, ref.time, rep.times = 100){
+       r.time <- system.time(for(i in 1:rep.times)
+                               res <- eval(expr))[1]/ref.time
+       names(r.time) <- NULL
+       return(list(res = res, r.time = r.time))
+  }
\end{Sinput}
\end{Schunk}

Comparing the relative timings we get the following result
(where timings are reported as percentages relative to
\code{convActuar}):
\begin{Schunk}
\begin{Sinput}
R> rep <- 100
R> refset <- speedref(quote(convActuar(N = 10, method = "lower")),
+                     rep.times = rep)
R> r1 <- speedcheck(expr = quote(convpow(D1 = D1, N = 10)),
+                   ref.time = refset$ref.time, rep.times = rep)
R> r2 <- speedcheck(expr = quote(convpow(D1 = Chisq(), N = 10)),
+                   ref.time = refset$ref.time, rep.times = rep)
R> r3 <- speedcheck(expr = quote(Chisq(df = 10)),
+                   ref.time = refset$ref.time, rep.times = rep)
R> res <- refset$res
R> D10 <- r1$res; Dex <- r2$res; Dcheck <- r3$res
R> round(refset$ref.time, 2)
\end{Sinput}
\begin{Soutput}
[1] 3.97
\end{Soutput}
\begin{Sinput}
R> print(round(c("actuar" = 1,"FFT" = r1$r.time,
+                "Chisq-Meth" = r2$r.time,
+                "exact" = r3$r.time)*100, 2))
\end{Sinput}
\begin{Soutput}
    actuar        FFT Chisq-Meth      exact
    100.00      39.29      24.69       7.05
\end{Soutput}
\end{Schunk}
As to accuracy, our algorithm still is competitive:

\begin{Schunk}
\begin{Sinput}
R> summary(abs(res$d(knots(res$d))[-c(1:8)] - p(D10)(res$x)))
\end{Sinput}
\begin{Soutput}
     Min.   1st Qu.    Median      Mean   3rd Qu.      Max.
0.0000000 0.0001000 0.0001000 0.0001927 0.0001000 0.0019000
\end{Soutput}
\begin{Sinput}
R> summary(abs(res$d(knots(res$d))[-c(1:8)] - p(Dex)(res$x)))
\end{Sinput}
\begin{Soutput}
     Min.   1st Qu.    Median      Mean   3rd Qu.      Max.
2.010e-08 1.000e-04 1.000e-04 2.455e-04 1.000e-04 3.120e-03
\end{Soutput}
\begin{Sinput}
R> summary(abs(p(Dex)(res$x) - p(D10)(res$x)))
\end{Sinput}
\begin{Soutput}
     Min.   1st Qu.    Median      Mean   3rd Qu.      Max.
0.000e+00 0.000e+00 0.000e+00 6.111e-05 1.758e-07 1.253e-03
\end{Soutput}
\end{Schunk}
Note that the computations with \code{aggregateDist} of \pkg{actuar}
get considerably more expensive if you pass to finer discretizations,
as we show in the following illustration which now cuts off
lower and upper $10^{-6}$-quantiles (instead of $10^{-5}$ beforehand)
and which uses $4$ times as many discretization points
(with only $30$ replications)---again we report
percentages relative to \code{convActuar}:
\begin{Schunk}
\begin{Sinput}
R> distroptions(TruncQuantile = 1e-6)
R> distroptions(DefaultNrFFTGridPointsExponent = 14)
R> rep <- 30
\end{Sinput}
\end{Schunk}
\begin{Schunk}
\begin{Sinput}
R> round(refset$ref.time,2)
\end{Sinput}
\begin{Soutput}
[1] 75.74
\end{Soutput}
\begin{Sinput}
R> print(round(c("actuar" = 1,"FFT" = r1$r.time,
+                "Chisq-Meth" = r2$r.time,
+                "exact" = r3$r.time)*100, 2))
\end{Sinput}
\begin{Soutput}
    actuar        FFT Chisq-Meth      exact
    100.00       1.82       0.37       0.13
\end{Soutput}
\begin{Sinput}
R> summary(abs(res$d(knots(res$d))[-c(1:8)] - p(D10)(res$x)))
\end{Sinput}
\begin{Soutput}
     Min.   1st Qu.    Median      Mean   3rd Qu.      Max.
0.000e+00 1.000e-05 1.000e-05 3.401e-05 1.000e-05 5.745e-04
\end{Soutput}
\begin{Sinput}
R> summary(abs(res$d(knots(res$d))[-c(1:8)] - p(Dex)(res$x)))
\end{Sinput}
\begin{Soutput}
     Min.   1st Qu.    Median      Mean   3rd Qu.      Max.
1.000e-10 1.000e-05 1.000e-05 4.172e-05 1.000e-05 7.835e-04
\end{Soutput}
\begin{Sinput}
R> summary(abs(p(Dex)(res$x) - p(D10)(res$x)))
\end{Sinput}
\begin{Soutput}
     Min.   1st Qu.    Median      Mean   3rd Qu.      Max.
0.000e+00 0.000e+00 0.000e+00 8.643e-06 1.560e-09 2.149e-04
\end{Soutput}
\end{Schunk}
%
\section{Conclusion}\label{conclusion}
With our implementation of a general default convolution algorithm
for distribtions in the object oriented framework of \proglang{R}
we provide a flexible framework which combines scalable accuracy
and reasonable computational efficiency. This framework lends itself
for introductory courses in statistics where students can easily sharpen
their intuition about how convolution and other arithmetic operations
work on distributions. It is however not limited to merely educational
purposes but can be fruitfully applied to many problems where one
needs the exact distributions of convolutions, as arising e.g.,
in finite sample risk of M estimators \citep{Ru:Ko:04},
actuarial sciences and risk management \citep{RS:10},
linguistics \citep{AL:12}, and Bingo premia calculations
\citep{K:R:11}.
%
\section*{Acknowledgement}
Both authors contributed equally to this work.
The first implementation of our FFT algorithm from end of 2003 is due to
our former student, T. Stabla, who also collaborated with us on this topic
until he left academia in 2006 and whom we warmly thank for his efforts.
We thank 
Prof.\ Gr{\"u}bel for drawing our attention to relevant literature on this
topic and two anonymous referees for their valuable comments.
Financial support from VW Foundation in the framework of project
\href{http://www.mathematik.uni-kl.de/~wwwfm/RobustRiskEstimation/}%
{``Robust Risk Estimation''} for which \pkg{distr} provides indispensable
infrastructure, is greatfully acknowledged.
\bibliography{litdb2} 

\end{document}